\documentclass[amssymb,pra,twocolumn,showpacs,footinbib]{revtex4}
\usepackage[]{graphicx,amsmath}
\usepackage{color}
\usepackage[all]{xy}
\usepackage{amsthm}

\def\id{{\mathchoice {\rm 1\mskip-4mu l} {\rm 1\mskip-4mu l} {\rm
1\mskip-4.5mu l} {\rm 1\mskip-5mu l}}}
\def\ket#1{| #1 \rangle}
\def\bra#1{\langle #1 |}

\def\ketbra#1#2{| #1 \rangle\!\langle #2 |}

\newcommand {\be} {\begin{eqnarray}}
\newcommand {\ee} {\end{eqnarray}}
\newcommand {\bp} {\begin{pmatrix}}
\newcommand {\ep} {\end{pmatrix}}
\newcommand{\ham}{\mathcal{\hat{H}}}
\newcommand{\X}{\hat{X}}
\newcommand{\Y}{\hat{Y}}
\newcommand{\Z}{\hat{Z}}
\newcommand{\U}{\hat{U}}
\newcommand{\A}{\hat{A}}

\newcommand{\E}{\mathcal{E}}

\begin{document}

\title{Using error correction to determine the noise model}

\author{Martin Laforest\footnote{email address : mlafores@iqc.ca}}
\author{Damien Simon}
\author{Jean-Christian Boileau}
\author{Jonathan Baugh}
\author{Michael J. Ditty}
\author{Raymond Laflamme}
\affiliation{Institute for Quantum Computing, University of Waterloo, Waterloo, ON, N2L 3G1, Canada.}

\date{\today}
\begin{abstract}\
Quantum error correcting codes have been shown to have the ability of making quantum information resilient against noise.  Here we show that we can use quantum error correcting codes as diagnostics to characterise noise. The experiment is based on a three-bit quantum error correcting code carried out on a three-qubit nuclear magnetic resonance (NMR) quantum information processor.  Utilizing both engineered and natural noise, the degree of correlations present in the noise affecting a two-qubit subsystem was determined. We measured a correlation factor of $c=0.5\pm0.2$ using the error correction protocol, and $c=0.3\pm0.2$ using a standard NMR technique based on coherence pathway selection.  Although the error correction method demands precise control, the results demonstrate that the required precision is achievable in the liquid-state NMR setting.

\end{abstract}


\maketitle

\section{Introduction}

The idea of using quantum mechanical systems as information processing devices was proposed more than two decades ago \cite{Fey82a}, and yet, experimental realization of such devices remains a challenge.  Ultimately, all physical realizations are faced with the presence of decoherence, or noise, caused by uncontrollable interactions with the environment \cite{Zur91a}.

To prevent the loss of coherence in the quantum mechanical processor, the theory of quantum error correction (QEC) has been developed \cite{Sho95a, Ste96a, BDSW96a, KL97a}.  QEC works by encoding the state of a system lying in a certain Hilbert space into a state in a larger Hilbert space.  The encoding is designed to make possible the recovery of the original information after noise has acted on the overall system, through decoding and syndrome measurement, as long as the the noise level falls below a certain threshold \cite{KLZ98a, KLZ98b}.  Many quantum error correction codes (QECC) have been developed for specific classes of noise models.  For example, there are codes that can correct arbitrary single qubit errors \cite{Sho95a, Ste96a, BDSW96a, LMPZ96a} but fail to correct multi-qubit errors.  This work shows how this failure can be used to extract information about the noise of the system.

Most QECC are developed for independent, or uncorrelated, error models, meaning that the errors happening on one qubit are assumed to be independent of the errors on other qubits.  Clearly, knowing whether or not there exist correlations in the noise model plays an  important role in choosing the best QECC for a given system.

The noise model can be determined exactly by performing process tomography \cite{CN97a, PCZ97a}. However, the number of experiments required for complete tomography grows exponentially in the number of qubits.  Often, process tomography is not needed and important (but partial) information about the noise can be extracted from fewer experiments.  Here, we demonstrate the use of a three-qubit QECC to extract the two-qubit noise correlation factor under a transverse relaxation process in nuclear magnetic resonance (NMR) (e.g. $T_2$ relaxation). 

Transverse relaxation is the main source of decoherence in liquid state NMR.  NMR was used to perform the first experimental implementation of QEC \cite{CMP+98a}, where it was shown that the three-qubit QECC could correct single-qubit phase flip errors caused by $T_2$ relaxation.   Here, we will first briefly review the fundamentals of NMR, then model the noise present in such systems and show how noise correlations can affect the fidelity of the QEC protocol. We will describe a series of natural and engineered noise experiments for determining the two-spin noise correlations present for the $^{13}$C subsystem of acetyl chloride (dissolved in deuterated chloroform). The experimental results are in agreement with those obtained using the standard NMR technique of coherence selection.  In light of the exquisite sensitivity of these experiments to control imperfections, the results also demonstrate the high degree of precision attainable in controlling nuclear spins in liquid state NMR. 

The results demonstrate that QEC can not only be used for correcting the effects of decoherence, but can also help to characterize the nature of those errors.  Moreover, as QEC is a requirement for scalable quantum information processing (QIP), this methodology is universal for probing noise correlations in physical systems suitable for QIP.

\section{NMR quantum computing}
Liquid state nuclear magnetic resonance has proven to be a useful system for experimentally benchmarking small-scale quantum information processing devices \cite{LKZ+97b, NKL98a, CMP+98a, KLMT00a, NMR+06a}. A NMR quantum information processor consists of an ensemble containing of order $10^{20}$ molecules with spin-$\frac{1}{2}$ nuclei dissolved in a liquid solvent. Placed in a strong homogeneous magnetic field, nuclear spins precess about the direction of the field, defined conventionally as the $z$-axis. The rate at which the spins precess is the Larmor frequency, and nuclei with distinct Larmor (resonance) frequencies can be mapped to qubits.  In the liquid state, picosecond-scale rotation and translation of the molecules causes spins on separate molecules to effectively decouple on the NMR timescale. Therefore, to a very good approximation, all molecules of the same type experience identical environments and the Hilbert space of the nuclear spin ensemble can be taken as that of a single molecule. Moreover, the rotational degree of freedom causes the internal dipolar interaction between the spins on each molecule to vanish.  At thermal equilibrium, the Boltzmann distribution gives a slight excess of spins pointing along the $+z$ direction, so that an average magnetization is present along $+z$. 

Control of the qubits is achieved by a radio-frequency (RF) Hamiltonian in which the frequency, phase and duration of the RF can be controlled externally.  Single-qubit rotations are performed using RF pulses resonant with the Larmor frequency of the targeted qubit.  By varying the RF duration and phase, rotations of arbitrary angle can be generated about any axis in the $xy$-plane. Two-qubit operations additionally use the natural coupling terms present in the internal Hamiltonian, which will be elaborated below.

\section{Hamiltonian and noise model}\label{noise}

The nuclear spin Hamiltonian in liquid-state NMR is composed of two types of terms, one corresponding to the single-spin Zeeman interaction (the term leading to precession) and bilinear terms corresponding to the scalar spin-spin coupling (J-coupling).  For a molecule with $N$ spin-1/2 nuclei, the weak coupling Hamiltonian is given by
\be
\ham&=&\frac{1}{2}\sum_{i=1}^{N}2\pi\nu_i\Z_i+\frac{\pi}{2}\sum_{i<j}J_{ij}\Z_i\Z_j
\ee
where $\nu_i$ are the Larmor frequencies, $J_{ij}$ is the coupling strength between spins $i$ and $j$, and $\Z_i$ is the $z$ Pauli matrix for spin $i$.  Note that when the condition $|\nu_i-\nu_j|>>J_{ij}/2$ does not hold (strong coupling regime), the scalar coupling operator takes the more general form $\vec{\sigma}_i\cdot\vec{\sigma}_j=\X_i\X_j+\Y_i\Y_j+\Z_i\Z_j$.

Despite the motional averaging that occurs in the liquid state, the $\Z\Z$ part of the intermolecular dipolar Hamiltonian is capable of creating relaxation, while the $\X\X+\Y\Y$ part still averages to zero due to the weak coupling approximation\cite{Lev01b}.  This $\Z\Z$ interaction couples with the molecular motion and give rise to rapidly fluctuating local magnetic field, which effectively presents itself as a variation of the Larmor frequencies.
This process is known as transversal relaxation.

Consider a single spin qubit surrounded by an environment $\mathcal{E}$ consisting of $N$ other spins -$\frac{1}{2}$. The dipolar coupling between the qubit and its environment  is described by the unitary evolution
\be
\U&=&\prod_{j\in\mathcal{E}}e^{-ib_j\Z\Z_j}
\ee
where $b_j$ is the interaction strength between the qubit and the $j^{th}$ spin of the environment for a certain amount of time.

The global system can be assumed to initially be in the state
\be
\rho_{glob}=\rho_{ini}\otimes\rho_{\mathcal{E}}
\ee
where $\rho_{ini}$ represents the initial state of the qubit and $\rho_{\E}$ is the state of the environment.  After the interaction with its environment, the final state of the qubit will be given by partial tracing the environment system.  Moreover, the interaction strengths have a certain distribution of value $q(\vec{b})$, so that the state affected by the noise have the form
\be\label{finalgeneral}
\rho_f&=&\int d\vec{b}q(\vec{b})\sum_{a\in\{0,1\}^N}\bra{a}\rho_{\E}\ket{a}e^{-i\xi_a\Z}\rho_{ini}e^{i\xi_a\Z},
\ee
where we have defined $\xi_a=\sum_{m}b_m(-1)^{a_m}$, $a_m$ being the $m^{th}$ digit of $a$.  In room temperature liquid state NMR, the deviation of the state of the environment from the completely mixed state is negligible, so that $\bra{a}\rho_{\E}\ket{a}=1/N$. Because the environment is isotropic, the distribution $q(\vec{b})$ is a symmetric function of the $b_j$'s.  Therefore, the summation over $a$ in Eq. \ref{finalgeneral} can be absorbed in a new distribution of $\vec{b}$ and by letting  $a=0^{\otimes N}$.

The final state is then represented by 
\be
\rho_f&=&\int d\alpha p(\alpha)e^{-i\alpha\Z}\rho_{ini}e^{i\alpha\Z}
\ee
where $\alpha=\sum_mb_m$ and $p(\alpha)$ is the distribution of $\alpha$ which takes into account the new distribution of the $b_j$'s.  The interaction of the qubit with the environment causes an incoherent averaging of $z$ rotations, which is equivalent to a variation of the Larmor frequency of the qubit. In liquid state NMR, $N$ is a large number and the central limit theorem indicates that $\alpha$ has a gaussian distribution. For a $M$ qubit system, this model generalizes to
\be
\rho_f&=&\int d\vec{\alpha}p(\vec{\alpha})e^{-i\vec{\alpha}\cdot\vec{\Z}}\rho_{ini}e^{i\vec{\alpha}\cdot\vec{\Z}}
\ee
where $\vec{\alpha}=(\alpha_1,\ldots,\alpha_M)$, $\vec{\Z}=(\Z_1,\ldots,\Z_M)$ and $p(\vec{\alpha})$ is the multivariate gaussian distribution \cite{Ton01a}
\be\label{multiplefinal}
p(\vec{\alpha})&=&\frac{1}{\sqrt{(2\pi)^M |\hat{\Sigma} |}}e^{-\frac{1}{2}\vec{\alpha}^T\cdot \hat{\Sigma}^{-1}\cdot\vec{\alpha}}.
\ee
$\hat{\Sigma}$ is the covariant matrix, or the correlation matrix, which takes the form
\be
\hat{\Sigma}_{ii}&=&\langle\alpha_i^2\rangle\nonumber\\
&=&\sigma_i^2\\
\hat{\Sigma}_{ij}&=&c_{ij}\sqrt{\sigma_i\sigma_j}
\ee  
where $\sigma_i^2$ is the variance of $\alpha_i$.  $c_{ij}$ is the correlation factor between $\alpha_i$ and $\alpha_j$, which has value
\be
c_{ij}&=&\frac{\langle\alpha_i\alpha_j\rangle}{\sqrt{\langle\alpha_i^2\rangle\langle\alpha_j^2\rangle}}.
\ee
From the Cauchy-Schwarz inequality,  $0\leq c_{ij}\leq 1$. For a single qubit, such a noise model will affect the state as
\be
\ketbra{k}{l}&\rightarrow&e^{\frac{-\sigma^2}{2}(1-\delta_{kl})}\ketbra{k}{l}
\ee
From empirical results of transverse relaxation in NMR, the state of a single spin decays in time as
\be
\ketbra{k}{l}&\rightarrow&e^{-(1-\delta_{kl})\gamma_2 t}\ketbra{k}{l}
\ee
where $1/\gamma_2=T_2$ is the relaxation time constant. The variance of the distribution of the interaction strength of a qubit with its environment can thus be related to its relaxation time constant by
\be
\sigma^2&=&2\gamma_2 t.
\ee

For two qubits, the noise correlation factor will affect the decay of their mutual state as follow: 
\be\label{twoqubitnoise}
\ketbra{km}{ln}&\rightarrow&
e^{-(1-\delta_{kl})\gamma_2^{(1)}t-(1-\delta_{mn})\gamma_2^{(2)}t-2c_{12}t\eta_{kl}\eta_{mn}\sqrt{\gamma_2^{(2)}\gamma_2^{(2)}}}\nonumber\\
&&\times\ketbra{km}{ln},
\ee
where $\eta_{ij}=\frac{1}{2}((-1)^i-(-1)^j)$.  If correlations in the noise affecting two qubits is present, the transverse relaxation will be faster for a two spin double quantum coherence (e.g. $\ketbra{00}{11}$ and $\ketbra{11}{00}$) and slower for a two spin zero quantum coherence (e.g. $\ketbra{01}{10}$ and $\ketbra{10}{01}$).

The correlation in the noise on two qubits can be understood  through distinguishability.  If two nuclei precess at the same Larmor frequency, they are magnetically equivalent and thus see the same environment.  The two spins will interact identically with the environment, thus yielding a correlation factor of 1.  Two spins of different nuclear specie are distinguishable and the environment will act differently on each of them. No correlation is expected the respective noise i.e. $c_{12}=0$.  If we consider two nuclei of the same species with slightly different Larmor frequency, they are distinguishable enough to perform independent control, but they are chemically ``near indistinguishable''. The effect of the environment is thus partially correlated, i.e $0<c_{12}<1$.

\section{Engineering the noise for two qubits}\label{engineer}
By explicitly expanding Eq. \ref{multiplefinal} for two qubits, the noise model takes a discrete Kraus form,
\be
\rho_f&=&\sum_i p_i\U_i\rho_{ini}\U_i^\dagger
\ee
 where the unitary Kraus operators $\U_i$ and their coefficients $p_i$ are given in Table \ref{Kraus}.  One can thus engineer the noise on two qubits with a series of six separate experiments, each of them implementing a different Kraus operator, and then adding the results with the corresponding coefficient.
\begin{table}[th!]
\begin{tabular}{|c|c|}
\hline\hline
$\U_i$ & $p_i$ \\
\hline
\hline
$\id$		&	$\frac{1}{4}\left(1+e^{-\gamma_2^{(1)}t}+e^{-\gamma_2^{(2)}t}+e^{-\gamma_2^{(1)}t-\gamma_2^{(2)}t-2c_{12}t\sqrt{\gamma_2^{(1)}\gamma_2^{(2)}}}\right)$\\
\hline
$\Z_1$		&	$\frac{1}{4}\left(1-e^{-\gamma_2^{(1)}t}+e^{-\gamma_2^{(2)}t}-e^{-\gamma_2^{(1)}t-\gamma_2^{(2)}t-2c_{12}t\sqrt{\gamma_2^{(1)}\gamma_2^{(2)}}}\right)$\\
\hline
$\Z_2$		&	$\frac{1}{4}\left(1+e^{-\gamma_2^{(1)}t}-e^{-\gamma_2^{(2)}t}-e^{-\gamma_2^{(1)}t-\gamma_2^{(2)}t-2c_{12}t\sqrt{\gamma_2^{(1)}\gamma_2^{(2)}}}\right)$\\
\hline
$\Z_1Z_2$		&	$\frac{1}{4}\left(1-e^{-\gamma_2^{(1)}t}-e^{-\gamma_2^{(2)}t}+e^{-\gamma_2^{(1)}t-\gamma_2^{(2)}t-2c_{12}t\sqrt{\gamma_2^{(1)}\gamma_2^{(2)}}}\right)$\\
\hline
$e^{-i\frac{\pi}{4}(\Z_1+\Z_2)}$		&	$\frac{1}{2}e^{-\gamma_2^{(1)}t-\gamma_2^{(2)}t}\sinh(2c_{12}t\sqrt{\gamma_2^{(1)}\gamma_2^{(2)}})$\\
\hline
$e^{i\frac{\pi}{4}(\Z_1+\Z_2)}$		&	$\frac{1}{2}e^{-\gamma_2^{(1)}t-\gamma_2^{(2)}t}\sinh(2c_{12}t\sqrt{\gamma_2^{(1)}\gamma_2^{(2)}})$\\
\hline
\hline
\end{tabular}
\caption{Kraus decomposition for the correlated noise on two qubits \label{Kraus}}
\end{table}
This Kraus decomposition demonstrates that the transversal relaxation in NMR is equivalent to a phase flip error, where the qubits undergo a phase flip given by the operator in the first column of Table \ref{Kraus} with a probability given by the second column. 

\section{Determining the correlation factor}
This section will explain how the noise correlation factor between two spins can be extracted using standard NMR techniques and how quantum error correction can be used to achieve similar results.  The details and results of the experimental implementation,  as well as a summary of the advantages of this new technique will then conclude this section.
\subsection{NMR techniques}
In NMR, measurement of transversal relaxation times ($T_2$'s) is a standard techniques and is implemented through single coherence decay and spin echo \cite{Han50a}.  The same technique is applicable to double coherences to extract the noise correlation factor between two spins.  Consider the following pulse sequence:
\be
\frac{\tau}{2}\rightarrow\pi_1\pi_2\rightarrow\frac{\tau}{2}.
\ee
where $\tau$ is a certain time delay and $\pi_i$ correspond to a $\pi$ pulse on nuclei $i$ around any axis in the $xy$-plane.  They are used to refocus the field inhomogeneities via spin echo. If we apply such  a pulse sequence to a state of the form
\be
\rho_{ini}=\ketbra{00}{11},
\ee
which can be created using standard NMR techniques of coherence selection such as phase cycling \cite{BKE84a} or field gradients, the noise model developed earlier predicts that the amplitude of such a state should decay as
\be
\ketbra{00}{11}&\rightarrow& e^{-\gamma_2^{(1)}t-\gamma_2^{(2)}t-2c_{12}t\sqrt{\gamma_2^{(1)}\gamma_2^{(2)}}}\ketbra{00}{11}.
\ee
In NMR, only single coherence state can be detected. A final $\frac{\pi}{2}$ pulse is thus needed on one of the spin to detect such a state.  By repeating the experiment for various value of $\tau$, one obtain a decay curve.  Once the values of $T_2$ are measured using single coherence decay experiments, it is possible to deduce the value of $c_{12}$ .

\subsection{Three qubit quantum error correction code}
\begin{figure}[t]
\includegraphics[scale=0.8]{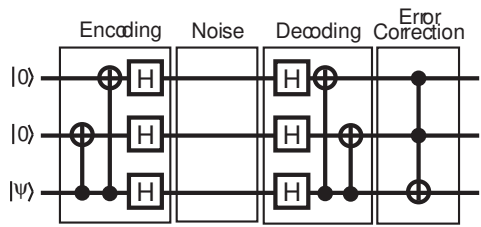}
\caption{The quantum circuit of the three qubit quantum error correction code.  The two qubit gates are C-NOTs and the three qubit gate is a Toffoli gate.}\label{QECcircuit}
\end{figure}
The three qubit quantum error correction code \cite{CMP+98a} can protect one qubit of information $\ket{\psi}$ against single qubit errors about one Pauli axis. The quantum circuit for this code can be found in Fig. \ref{QECcircuit}.  If errors happen during the noise period, it can be shown that this code corrects any single qubit phase error, i.e. errors of the form $\Z_1, \Z_2$ or $\Z_3$, but fails at correcting multiple phase errors, i.e. $\Z_1\Z_2, \Z_1\Z_3, \Z_2\Z_3$ and $\Z_1\Z_2\Z_3$. 

As seen above, the natural noise present in NMR consists of a phase flip. A valid measure to quantify the effect of the noise on the system is to consider the fidelity of entanglement $F_E$ \cite{Sch96a}, which corresponds to averaging the state-correlation for the density matrix states $\X$, $\Y$ and $\Z$.  In other words, the state-correlation $f_{\A}$ for an initial state $\A$ consist on the amount of polarization in the output relative to the input.  The fidelity of entanglement is then given by
\be
F_E&=&\frac{1}{4}(1+f_{\X} +f_{\Y} +f_{\Z}).
\ee
 If the decoherence is caused solely by the transversal relaxation, the fidelity of entanglement over time of such a protocol is given by
\be\label{fidQEC}
F_E&=&\frac{1}{4}\left[2+e^{-\gamma_2^{(1)}t}+e^{-\gamma_2^{(2)}t}+e^{-\gamma_2^{(3)}t}\right.\nonumber\\
&&\left.-e^{-\gamma_2^{(1)}t-\gamma_2^{(2)}t-\gamma_2^{(3)}t}\cosh(2c_{12}t\sqrt{\gamma_2^{(1)}\gamma_2^{(2)}})\right],
\ee
where it has been assumed that the noise affecting qubit 3 was uncorrelated with the other qubits (because qubit 3 is represented by a separate nuclear species in our experiment). The correlation factor can be extracted from the deviation of the fidelity from unity, due to the failure of the code.

As demonstrated in Sec. \ref{engineer}, it is possible to engineer the correlated noise on two spins using six different experiments.  If we want to engineer the noise for a third uncorrelated qubit, it is done with twelve experiments, using the union of two sets of Kraus operators/coefficients given by
\be\label{Kraus3}
\{\U'_k,p'_k\}&=&\{\U^{1,2}_i,(1-q)p_i^{1,2}\}\bigcup\{\U^{1,2}_j\Z_3,qp_j^{1,2}\}
\ee
for $k=1\ldots12$ and $i,j=1\dots6$ and where $q=\frac{1}{2}(1-e^{\gamma_2^{(3)}t})$ corresponds to the probability of the uncorrelated qubit to undergo a phase flip and $\U^{1,2}_i$ and $p_i^{1,2}$ are the correlated noise Kraus operators/coefficients given in Table \ref{Kraus}.

Therefore, we can implement the QECC using those twelve noise operators and obtain the fidelity decay for various value of $c_{12}$ and $t$ .

\subsection{The experiment}
\begin{figure}[ht!]
\includegraphics[scale=0.7]{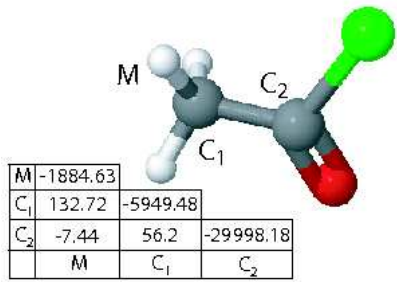}
\caption{The $^{13}C$-labeled acetyl chloride molecule.  The diagonal elements of the table gives the chemical shift (difference in Larmor frequency) for each nucleus with respect to a reference frequency (700.13 MHz for the hydrogens and 176.05 MHz for the carbons).  The three hydrogens forming the methyl group are indistinguishable and form a spin-$\frac{1}{2}$ and $\frac{3}{2}$ subspace. The spin-$\frac{1}{2}$ subspace was selected using a three-step pulse sequence and ``crusher'' field gradients \cite{CPH98a}.\label{acetyl}}
\end{figure}

The theory laid down in the previous section assumed that the system is composed of two noise correlated qubits and one uncorrelated qubit. As seen in section \ref{noise}, such a system can be found in a molecule containing two spins of the same species with different Larmor frequency and one of a different kind.  For this experiment, we have chosen the $^{13}C$-labeled acetyl chloride dissolved in deuterated chloroform and used a 700 MHz Bruker Avance NMR spectrometer with dual inverse cryoprobe.  The structure, chemical shifts and J-coupling strengths of the molecule are given in Fig. \ref{acetyl}.  For this molecule, the assumption of weak coupling used throughout section \ref{noise} is fulfilled due to the large chemical shift difference between the two carbons.

\begin{table}[b]
\begin{tabular}{|c|c|c|}
\hline\hline
Nucleus & $T_1$&$T_2$\\
\hline\hline
$M$ & $4.0\pm0.1$&$1.2\pm 0.1$\\
\hline
$C_1$ & $7.9\pm0.4$ & $2.1 \pm 0.1$\\
\hline
$C_2$ & $15.2\pm0.8$ & $0.24 \pm 0.03$\\
\hline\hline
\end{tabular}
\caption{$T_1$ and $T_2$ values for the acetyl chloride.  It can be seen that for the maximal duration of the experiment ($\sim 300\,ms$), the effect caused by $T_1$ relaxation can be neglected.\label{T2}}
\end{table}

The $T_2$'s for each nuclei have been determined using a series of spin echo experiments for various delays and their values are given in Table \ref{T2}.
To implement the quantum error correction code on this molecule, the circuit in Fig.
\ref{QECcircuit} was first converted into gates implementable in NMR, which consist of single qubit rotations about any axis in the $xy$-plane or around the z-axis,  and J-coupling evolutions. A J-coupling of length $\tau=\frac{1}{2J}$ is locally equivalent to a C-NOT.  Moreover, the z-rotations can be done instantaneously by changing  the phase of subsequent pulses. This ideal NMR pulse sequence was then fed into a homemade compiler  which estimates the first order phase and coupling errors during the pulses and then tracks the phase of the subsequent pulses and optimize the refocusing scheme and J-coupling delays to minimize overall coupling errors \cite{KLMT00a}.  Spatial averaging \cite{KLMT00a} was used to initialized the states $\ketbra{00}{00}_{12}\otimes X_3, \ketbra{00}{00}_{12}\otimes Y_3$ and $\ketbra{00}{00}_{12}\otimes Z_3$, from the thermal state of a liquid state NMR system.

The quantum error correction code was first implemented using engineered errors in twelve experiments. The purpose of analyzing engineered noise is to be able to generate different fidelity of entanglement decay curves corresponding to different value of correlation factors.   It is done by adding each experiments weighted by the corresponding coefficient given in Table \ref{Kraus}. Once a fidelity decay curve is obtained for the natural noise, it is possible to extract the correlation factor by comparing to which engineered noise fidelity decay curve the natural noise curve correspond to.

The natural noise fidelity decay curve was obtained by implementing the identity map  during the noise section of Fig. \ref{QECcircuit}. To perform this implementation the spins could not be simply decoupled from one another using multiple $\pi$ pulses (e.g. the Hadamard refocusing scheme \cite{JK99a}).  Under such a refocusing scheme for a time $t$, the double coherence terms in a density matrix spend as much time in zero quantum coherence as in double quantum coherence.  Therefore, from Eq. \ref{twoqubitnoise}, the correlation factor term in the exponential decay cancel and do not affect the decay of the double coherence term.  

If we let the natural noise act on the system for a period $\tau=\frac{n}{J_{C_1C_2}},\,n\in\mathbb{N}$,  the overall evolution is an identity and the terms of the density matrix containing a double coherence for the two carbons  have remained in double coherence during the entire delay.  The field inhomogeneities can be refocused by applying simultaneous $\pi$ pulses on the carbons which leaves the J-coupling evolution untouched.
\begin{figure}[ht!]
\includegraphics[scale=0.4]{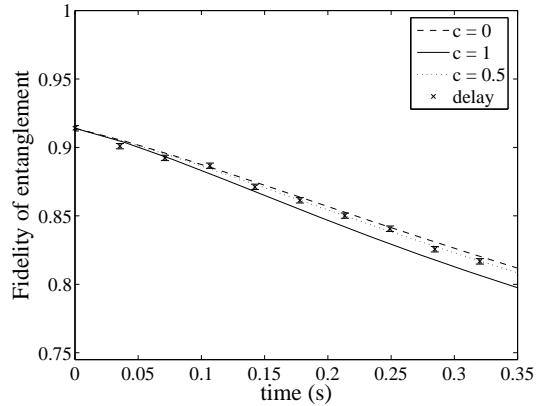}
\caption{Experimental results.  The lines correspond to the fidelity decay for noise correlation factors of 0, 0.5 and 1 as a function of time simulated from the measured $T_2$'s and the experimental fidelities obtained by implementing engineered noise. The points are the fidelities when the system is affected by natural noise for a various amount of time.\label{results}}
\end{figure}

The fidelities of our experiments have been extracted by fitting every peak of the NMR spectra using Lorentzian shape curves.  The resulting values can be seen in Fig. \ref{results}, where the curves for the fidelity of entanglement for engineered noise are shown for correlation factors of 0, 0.5 and 1.  Ten values of the fidelity decay obtained by applying natural noise are shown, from which we extracted a correlation factor of $c_{12}=0.5\pm0.2$.  

This experiment needs a high degree of precision, since within the time interval used to implement natural noise (from 0 to 320 ms),  the maximum difference between the $c_{12}=0$ and $c_{12}=1$ curves is 3$\%$.  Integrating the square of the noise of a spectrum over a region corresponding to the width of a signal peak estimates the signal to noise ratio to be of the order of $1\%$.  Therefore, the fluctuation of the measured fidelities due to the noise explains the large uncertainty on the measured correlation factor. 

Using the usual NMR technique of double coherence decay, the noise correlation factor between $C_1$ and $C_2$ was determinded to be $c_{12}=0.3\pm0.2$.  The interval using QECC agrees with the value obtained using the traditional double-quantum coherence decay technique, to within experimental error.

\subsection{Discussion}

By comparing the above two techniques to extract the noise correlation factor between two spins, one could  argue that the QECC technique is much more involved then the standard NMR technique, while yielding to the same conclusion.  The goal of the present experiment was to demonstrate that the use of QEC to probe the noise present in a system is feasible and that the control necessary to get the error information is achievable.  From there, it is possible to generalize this technique to any physical system with more complex noise model.  If the noise contains not only phase errors, but also bit errors and/or a combination of the two, the same technique could be applied using more complex QECC, such as the five qubit code \cite{LMPZ96a}.

Other technical advantages arise from the signal detection.  Using the NMR technique, there is an doubly exponential decay in the signal amplitude for a double coherence decay (see Eq. \ref{twoqubitnoise}).  From the nature of the QECC technique, the signal decays slower, thus allowing better statistics and analysis. If the system under analysis contains three spins of the same type, there would be a possibility of three different correlation factors $c_{12}$, $c_{23}$ and $c_{13}$.  Using standard NMR technique, three different experiments with different initial states would be needed to extract those three values.  Using the QECC technique, only the noise portion of the pulse sequence would need adjustment by changing the refocusing scheme in order to refocus the unwanted correlation,e.g. a $\pi$ pulse on qubit 1 would cancel the correlation $c_{12}$ and $c_{13}$ for the reasons explained earlier.  

Finally, this technique could be used to validate our assumption that the noise is effectively gaussian.  In the case where the system contained three noise correlated spins, our gaussian assumption ensures that the noise is only pairwise correlated, i.e. $c_{123}=0$.  If it is so, a triple coherence decay curve should be described using only the $T_2$ values and the pairwise correlation factors. This curve can be obtain using the NMR technique of triple coherence decay, but would yield a curve that decays triply exponentially.  On the other hand, a curve affected by a triple correlation factor could be obtained by the same QECC pulse sequence by letting all the noise correlations act during the noise part of the pulse sequence.  In the case where that curve would not be described properly using only the pairwise correlations, it would be an indication of the failure of the noise model.

\section{Conclusion}
In this work, we have demonstrated that QEC can be used to probe a physical system and extract partial, but important information about the noise model without having to perform full quantum process tomography.  The technique was implemented successfully in a liquid-state NMR quantum information processor, but is applicable to any QIP device in which standard quantum error correction can be carried out.

\begin{acknowledgments}
This work has been supported by NSERC, MITACS,  the National Security Agency (NSA) under Army Research Office (ARO) contract number W911NF-05-1-0469 and CIAR.
\end{acknowledgments}

\end{document}